\documentclass[%
reprint,
 amsmath,amssymb,
 aps,
]{revtex4-2}

\usepackage{graphicx}
\usepackage{dcolumn}
\usepackage{bm}
\usepackage{color}
\usepackage{float}

\usepackage{subfigure,lipsum}
\newcommand{\hy}[1]{\textcolor{black}{{#1}}}
\newcommand{\zl}[1]{\textcolor{black}{{#1}}}

\begin{document}

\preprint{APS/123-QED}

\title{2D numerical simulation of lunar response to gravitational waves using finite element method}

\author{Lei Zhang$^{1}$}
  \thanks{The first two authors contributed equally to this work.}
\author{Han Yan$^{2,3}$}
  \thanks{The first two authors contributed equally to this work.}
\author{Xian Chen$^{2,3}$}
  \email{xian.chen@pku.edu.cn}
\author{Jinhai Zhang$^{1}$} 
  \email{zjh@mail.iggcas.ac.cn}

\affiliation{$^{1}$Research Center for Deep Earth Technology and Equipment, Institute of Geology and Geophysics, Chinese Academy of Sciences, Beijing 100029, China}

\affiliation{$^{2}$Department of Astronomy, School of Physics, Peking University, 100871 Beijing, China}

\affiliation{$^{3}$Kavli Institute for Astronomy and Astrophysics, Peking University, 100871 Beijing, China}

\date{\today}

\begin{abstract}

The response of the Moon to gravitational waves have been \zl{previously} carried out using analytical or semi-analytical models assuming ideal lunar
structures. Such models are advantageous for their high-speed calculation but
fail to account for the extremely heterogeneous subsurface and/or interior
structures of the Moon. Numerical calculations are needed, but it is
challenging to model the topography and lateral heterogeneity of the Moon. In addition, the computational cost is great \zl{because GW modeling usually requires long-physical-time simulations}.  As a first step towards overcoming the above difficulties, we employ a two-dimensional \zl{high-order finite element method (spectral element method)} to numerically simulate the lunar response to gravitational waves.  We verify our method by comparing our numerical results with those semi-analytical solutions. Based on such comparison, we also analyze the limitation of the two-dimensional simulation.  Our work breaks a new way towards the precise simulation of realistic lunar response to gravitational waves in the future and lays down a solid foundation for three-dimensional numerical simulations.  
\end{abstract}
                           
\maketitle


\section{\label{sec:intro}Introduction }

The search for gravitational wave (GW) in both $10-10^2$ Hz
\cite{2016PhRvL.116f1102A} range as well as nano-Hz
\cite{2023RAA....23g5024X,2023A&A...678A..50E,2023ApJ...951L...9A,2023ApJ...951L...6R}
range has achieved great success, which strongly motivates the
search in other frequency bands.  For example, several space-based
interferometers have been proposed to detect milli-Hertz (mHz) GWs, including
the Laser Interferometer Space Antenna (LISA \cite{2017arXiv170200786A}),
TianQin \cite{2016CQGra..33c5010L}, and Taiji \cite{2021PTEP.2021eA108L}. For
the deci-Hertz ($0.1$ Hz, or deci-Hz) band, one possibility which has been
considered for a long time is to use Earth or the Moon as a resonator to detect GWs
\cite{1960PhRv..117..306W,1968PhT....21d..34W,1974ApJ...187L..49M}. In this
direction, there is increasing interest recently to use an array of lunar
seismometers to increase the sensitivity to GW, such as the proposed Lunar GW
Antenna (LGWA)
\cite{2021ApJ...910....1H,2023SSRv..219...67B,2024arXiv240409181A} and other
similar proposals \cite{2023SCPMA..6609513L}.

Several recent works carefully looked into the response functions of the Moon
to GWs
\cite{2019PhRvD.100d4048M,PhysRevD.109.064092,2024JCAP...07..028B,2024arXiv241109559M}.
They resolved the apparent discrepancy between the results derived using
different forms of the GW force density, and provided a consistent
understanding of how to apply the response functions. However, these works also
showed that the result is sensitive to (i) the details in the lunar structure,
such as the fine \zl{subsurface} structure or the parameters of different \zl{interior layers}
\cite{2021ApJ...910....1H,2023SCPMA..6609513L,2024PhRvD.110f4025B,2024arXiv241109559M},
as well as (ii) the details in the numerical calculation of the normal modes,
e.g., assuming spherical Moon or infinite half plane, using MINEOS or other
eigen-mode solvers \cite{2024PhRvD.110f4025B,2024PhRvD.110f4034K}.  To help us
better decide the instrument sensitivity and select landing site for lunar seismic GW detector, a global response function accounting for realistic lunar
structure is still essential. 

As we know, the Moon had experienced long-term and heavy bombardments in history, which resulted in a great number of fractures and holes in the lunar
crust \cite{lognonne2003new,garcia2011very,weber2011seismic,nunn2020lunar,2020Lunar,zhang2022strong}.
These fine structures could distort the path of seismic wave propagation
\cite{zhang2022strong}. Besides, the topography  of the Moon is fairly rough,
due to the impacts and volcano activities \cite{wang2024returned,zhang2024lunar}. However, previous theoretical approaches or semi-analytical methods have difficulty in resolving the topography and lateral heterogeneity of the Moon. Consequently, it is necessary to employ purely numerical simulation techniques to establish a nonuniform lunar model and calculate its response to GW. Among the various numerical simulation
methods available, the finite element method (FEM) is widely utilized due to
its flexibility in handling geometric shapes \cite{Nicolas1999A}, coupling of
multiple physical fields\cite{zhang2023thermal}, and well-developed parallel
computing \cite{yagawa1993parallel}. Therefore, this study adopts the spectral
element method (SEM) \cite{komatitsch2005spectral}—a high-order FEM—to model
the entire Moon and calculate its seismic response to GWs.

The paper is organized as follows. In Section~\ref{sec:theo} we revisit the
theory of calculating the lunar response functions, and provide an updated
formula in the frequency domain. In Section~\ref{sec:NumSet}, we illustrate the
settings and related details in our FEM simulation. In
Section~\ref{sec:DataAnaly}, we present our new results from FEM
calculation, focusing on the angular dependency as well as the radial and
horizontal response functions. In Section~\ref{sec:diss}, we discuss the
effects which can affect the accuracy of our simulation, including the angular
resolution and the 2D nature of our model. In Section~\ref{sec:ccl}, we draw a conclusion to this work, and point out several caveats to be investigated in the future. Throughout the paper we adopt the International System of Units, unless mentioned otherwise. 

\section{\label{sec:theo}Review of the theory of lunar response to GW}

In this section, we first review the basic formulation of the lunar response to
GWs. We only modify the part related to the source-time function (STF), 
which allows us to derive the response function for a wide range of frequencies
at once, in a single simulation. \hy{As a result, we will omit some details of the formulae, which can all be found in our references below, and focus on the modification compared with previous formulation.}

As it has been studied in many previous works \cite{1983NCimC...6...49B,2019PhRvD.100d4048M,PhysRevD.109.064092,2024arXiv241109559M}, \hy{for a spherically-symmetric radially-heterogeneous sphere, its response to GWs can be obtained using Green's function method.} Given a linearly polarized GW propagating along $z-$axis with an amplitude $h_{0}$ and a STF $g(t)$,  
\begin{eqnarray}
    \mathbf{h} && =   h_{0} \epsilon _{ij} g(t)  ~,  \nonumber \\ \epsilon _{ij} && = \begin{bmatrix}
 1 & 1 & 0\\
 1 & -1 & 0\\
 0 & 0 & 0
	\end{bmatrix} ~, \label{eq:GWtensor}
\end{eqnarray}
the surface response of a radially heterogeneous elastic sphere can be written as: 
\begin{eqnarray}
    \vec{\xi }  \left ( \theta ,\varphi ,t \right ) = h_{0} \sum_{n} \sum_{m = -2}^{2} 
 \vec{s } _{nm} \left ( \theta ,\varphi  \right ) \Re \left \{ \bar{g}_{n}\left ( t \right ) \right \}   f^{m} \alpha _{2n}   \label{eq:surfsol}
\end{eqnarray}
where $(\theta,\varphi)$ follow the standard definition in a spherical coordinate, and the corresponding location in $(x,y,z)$ coordinate at the surface is $R\left ( \sin \theta \cos \varphi ,\sin \theta \sin \varphi, \cos \theta  \right )$ ($R$ is the lunar radius). $\vec{s } _{nm}$ is the displacement eigenfunction\hy{ constructed by the spherical harmonics and radial eigen-functions, depending only on the structure of the Moon}. $f^{m}$ is the ``pattern function'' related to GW wave vector and polarization, \hy{and hence is constant for the fixed GW tensor in Eq.~(\ref{eq:GWtensor})}. $\alpha _{2n}$ is the part depending on the radial structure \hy{as well as the external force density}, and $\bar{g}_{n}\left ( t \right )$ is the time-dependent part we will discuss below. \hy{Eq.~(\ref{eq:surfsol}) can be applied to different types of external force density, including Dyson-type force \cite{2019PhRvD.100d4048M,PhysRevD.109.064092} and tidal force \cite{PhysRevD.109.064092,2024arXiv241109559M}, which leads to the modification of $\alpha_{2n}$. The form of the force density we use in this paper is discussed with more details in Sec.~\ref{sec:NumSet}.}

According to previous analyses \cite{1983NCimC...6...49B,2019PhRvD.100d4048M,2024arXiv241109559M}, in the limit of $Q_{n} \gg 1$, the time-dependent part $\bar{g}_{n}\left ( t \right ) $, can be calculated as:
\begin{eqnarray}
    \bar{g}_{n}\left ( t \right ) =  \frac{1}{2\pi } \int \mathrm{d}\omega e^{i\omega t} \frac{1+\frac{i\omega }{2Q_{n} \omega _{n} }  }{\omega _{n}^{2} - \omega ^{2} + i\omega _{n}\omega /Q_{n} } g (\omega ) ~, \label{eq:stf}
\end{eqnarray}
where \hy{$\omega = 2\pi f$, }and $g (\omega ) =  \int \mathrm{d}t e^{i\omega t} g(t)  $ is the Fourier transform of the STF. \hy{$\omega_{n}$ and $Q_{n}$ are the $n$-th eigen-frequency and quality factor in spheroidal mode with $l=2$ respectively, which can be calculated by MINEOS package \cite{1988MINEOS}. Their values in this work are almost the same as in Ref.~\cite{PhysRevD.109.064092}, while the slight deviation is simply due to a coarser lunar model used in FEM simulations.} Thus following the definitions in \cite{PhysRevD.109.064092,2024arXiv241109559M}, the lunar response per unit strain can then be written in a more concise form in the frequency domain:
\begin{eqnarray}
    \vec{\xi } \left ( \theta ,\varphi ,f \right )/h_{0}  && = \mathcal{F} \left \{ \vec{\xi } \left ( \theta ,\varphi ,t \right ) \right \} /h_{0} \nonumber \\&&  = g(\omega  )\times \nonumber\\ &&\Bigg[ T_{r}(\omega  ) \sum_{m}  f^{m}   \mathcal{Y}_{2m} \left ( \theta ,\varphi  \right ) \hat{e}_{r} \nonumber \\ && + T_{h}(\omega  ) \sum_{m}  f^{m} \partial _{\theta }  \mathcal{Y}_{2m} \left ( \theta ,\varphi  \right ) \hat{e}_{\theta }  \nonumber \\
    &&  +  T_{h}(\omega  ) \sum_{m}  f^{m}  \frac{\partial _{\varphi  }  \mathcal{Y}_{2m} \left ( \theta ,\varphi  \right )}{\sin \theta } \hat{e}_{\varphi   } \Bigg] ~, \label{eq:sol_perh}
\end{eqnarray}
in which \hy{$\mathcal{Y}_{2m}$ is the real spherical harmonics,} and the radial and horizontal response functions in the frequency domain are defined as
\begin{eqnarray}
    && T_{r}(\omega  ) \equiv \sum_{n} U_{2n} \alpha _{2n}\Re \left \{ \frac{1+\frac{i\omega }{2Q_{n} \omega _{n} }  }{\omega _{n}^{2} - \omega ^{2} + i\omega _{n}\omega /Q_{n} }  \right \} \nonumber \\ && T_{h }(\omega  ) \equiv \sum_{n} \frac{V_{2n}}{\sqrt{6} }  \alpha _{2n}\Re \left \{ \frac{1+\frac{i\omega }{2Q_{n} \omega _{n} }  }{\omega _{n}^{2} - \omega ^{2} + i\omega _{n}\omega /Q_{n} }  \right \} ~,
\end{eqnarray}
\hy{where $U_{2n}$, $V_{2n}$ are two types of radial eigen-functions depending only on the structure. In our previous work \cite{PhysRevD.109.064092}, we use MINEOS package to calculate $\omega_{n}$, $Q_{n}$, $U_{2n}$, $V_{2n}$, and then obtain the value of $\alpha_{2n}$ for a certain form of external force density. The absolute values of $T_{r}$ and $T_{h}$ in this work should then be close to previous results from MINEOS approach, at least in the low frequency region. We will examine this issue by FEM simulation.}

In the numerical simulations of this work, \hy{we only consider a 2D model of the moon, which might be similar to the response of the sphere} on the $x-y$ plane (i.e., $\theta = \pi/2$). In this case, Eq.~(\ref{eq:sol_perh}) can be further simplified as
\begin{eqnarray}
    \vec{\xi } \left ( \varphi ,f \right )/h_{0}  &&   = g(\omega  )\times \nonumber\\ &&\Bigg[ T_{r}(\omega  ) \left ( \sin 2\varphi + \cos 2\varphi  \right ) \hat{e}_{r} \nonumber  \\
    &&  + 2 T_{h}(\omega  ) \left ( -\sin 2\varphi + \cos 2\varphi  \right ) \hat{e}_{\varphi   } \Bigg] ~. \label{eq:sol_perh_simp}
\end{eqnarray}
\hy{We will discuss the validity and deficiency of this 2D simplification in Sec.~\ref{sec:diss} and Appendix \ref{app:SpheCylin}.}

\section{\label{sec:NumSet}Finite element simulation}

\begin{figure*}[ht]
    \centering
    \includegraphics[width=16cm]{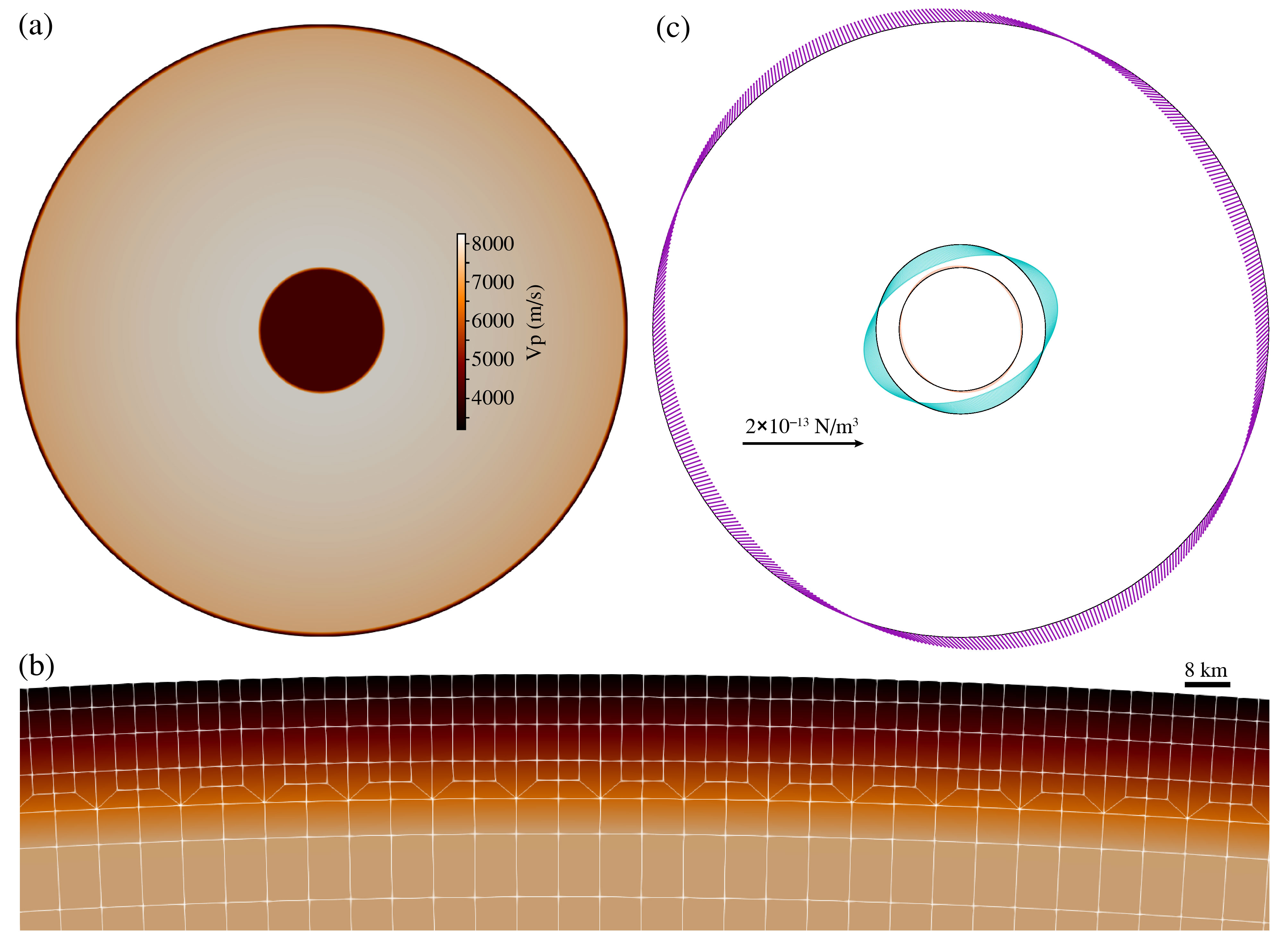}
    \caption{The SEM model for our simulation. (a) Layered global model of the entire Moon. The color represents the value of the P-wave velocity; (b) Enlarged regional crust grids near the azimuth of $0^{\circ} $, with a surface element size of about $3.7$ km. (c) The distribution of the 
force density vector with an azimuthal resolution of $1^{\circ} $. The black arrow 
shows a scale of $2\times10^{-13}$ N/m$^{2}$ for reference.}
    \label{fig:model}
\end{figure*}

Although theoretical or semi-analytical solutions can provide the response of lunar GW, they do not account for the rough topography (e.g., impact craters or volcanos) or the fluctuant subsurface structures (e.g., local geological \hy{bodies} and asymmetric lunar core). Therefore, there is an urgent need to develop a practicable scheme of numerically simulating the lunar response to GWs. In this section, we describe our scheme of 2D FEM simulation and the corresponding parameter configurations.


The SEM has been widely used in the field of seismic wave simulations in the
past decades (\cite{zhang2020procedure}) due to its high precision and great
feasibility in high performance parallel computing
(\cite{komatitsch2005spectral}). Herein we extend the 2D SEM to 
study the impact of GW on the Moon. Our simulation is based on
SPECFEM2D (\cite{komatitsch1998spectral,komatitsch1999introduction}), 
and we implement the forces induced by
GWs in the program and conduct a long-time simulation. 
A global 2D lunar model composed of 140,576 fourth-order
spectral elements has been built up (Fig.~\ref{fig:model}a), based on the same
model as in Ref.~\cite{PhysRevD.109.064092} with coarser layering near the
surface. The original model file \textit{Model\_FEM.txt} can be found in
\cite{webs}. The grid size is approximately 3.7 km in the horizontal direction
on the ground surface, and it increases according to the seismic wave velocities to avoid the artifacts caused by numerical dispersion that is associated with using course grid, as shown in Fig. \ref{fig:model}b. The highest frequency that can be accurately resolved by the model is approximately 0.2 Hz. The maximum allowed time step is approximately 0.035~s from the perspective of numerical stability. 


Based on our previous work \cite{PhysRevD.109.064092}, we decide to choose a Dyson-type force density \cite{1969ApJ...156..529D}, 
\begin{eqnarray}
    \vec{f} = \nabla \mu \cdot \mathbf{h} ~, \label{eq:forcedens}
\end{eqnarray}
to calculate the lunar response to GWs, where $\mathbf{h}$ has the same form as
in Eq.~(\ref{eq:GWtensor}).  This kind of force density is mainly located in
the layers with large radial variation of shear modulus $\mu ~(=\rho v_s^2)$,
so it should be varied correspondingly when the lunar model (thereby, grids)
near these layers changes, in order to maintain the total value of the force.
The distribution of the force density is shown in Fig.~\ref{fig:model}c, where
we have assumed $h_0 = 10^{-20}$. Here we add force vectors directly.  The
magnitude of the force equals force density times the volume involved, where
the \zl{volume involved} should be calculated by the force-covering area times
$1$~m, the thickness of our 2-D Moon model \cite{zhang2022dichotomy}. The
spatial resolution in the tangential direction (azimuth resolution) is
$1.00^{\circ}$ (Fig.~\ref{fig:model}c). As a result, $4320$ individual forces
have been applied in the entire Moon. 

\begin{figure}[htp]
    \centering
    \includegraphics[width=8cm]{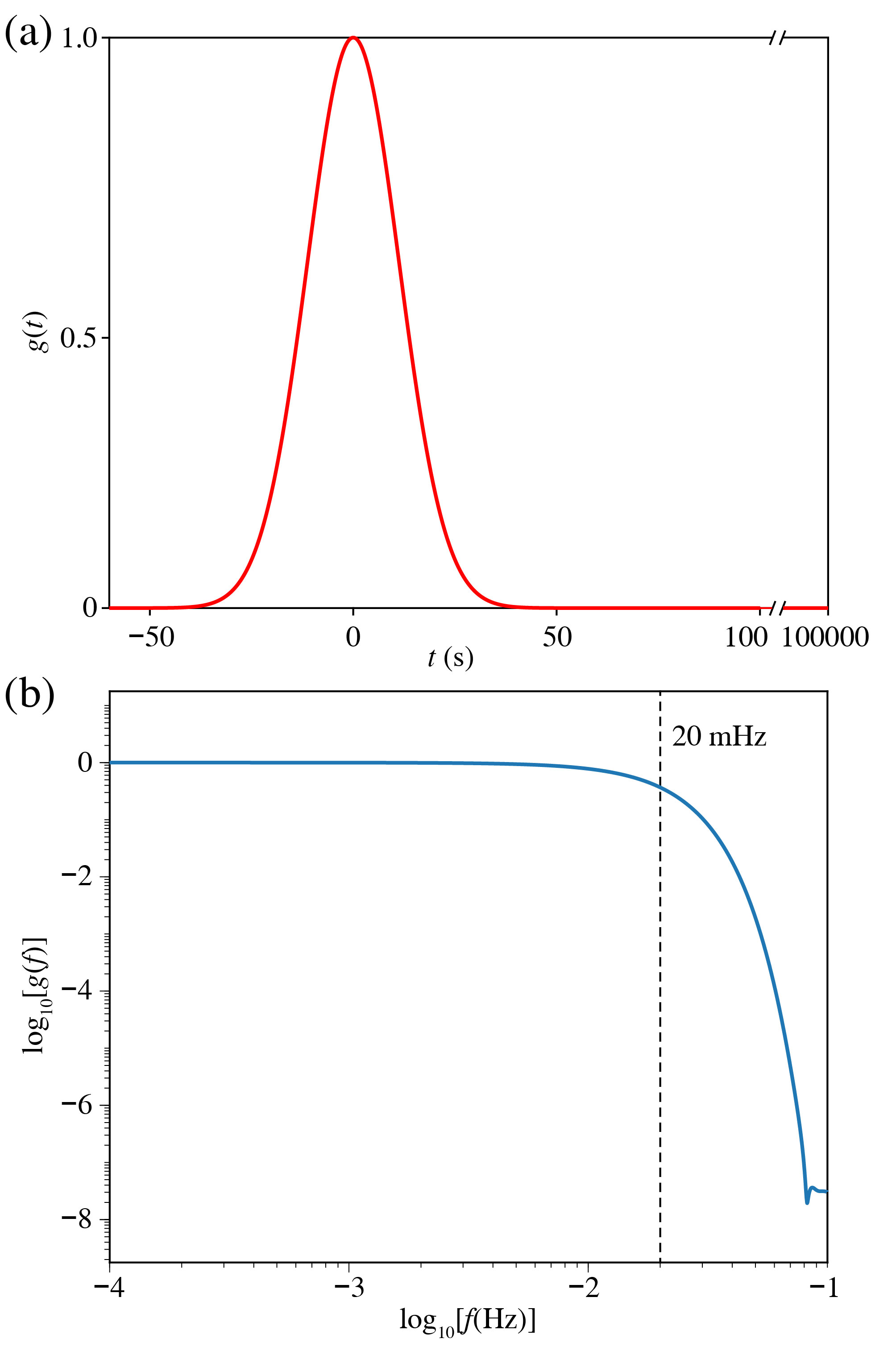}
    \caption{\hy{STF} (dimensionless) with a dominated frequency of $20$ mHz. 
	The result is normalized to have a maximum value of $1$. (a) The STF $g(t)$ for  
	a duration of 100,000~s. (b) The normalized amplitude spectral distribution $g(f)$.}
    \label{fig:stf}
\end{figure}

Each force can be regarded as an individual point force. For the purpose of our
numerical simulations, the ideal STF should result in a perturbation of the
Moon aross a wide frequency band since we focus on a wide frequency range of GW
(1$\sim$20 mHz). Herein we use the STF of Gaussian wavelet with a dominant
frequency of 20 mHz. The entire simulation duration is up to 100,000~s,
which allows seismic waves to propagate inside the Moon for tens of
rounds. The spectrum of the STF is presented in Fig.~\ref{fig:stf}. We can see
that the normalized amplitude is flat in the frequency range $<$ 20 mHz. The
simulation was completed on the National Supercomputing Center in Wuxi, China.
An individual node of the cluster has a total memory of 512 GB and consists of
64 cores, which means that up to 64 processes can simultaneously share the
memory.  Since we conducted a long-time simulation to study the response of the
Moon over a period of 100,000~s, \hy{a} large amount of memory is always required even for a 2D case. Therefore, we restrict the number of \hy{tasks} on each node to 10 to ensure that each \hy{task} has sufficient memory. As a result, the wall-clock time consumption is about 0.5 hours using 192 nodes with 12,288 cores.

\begin{figure*}[htp]
    \centering
    \includegraphics[width=14cm]{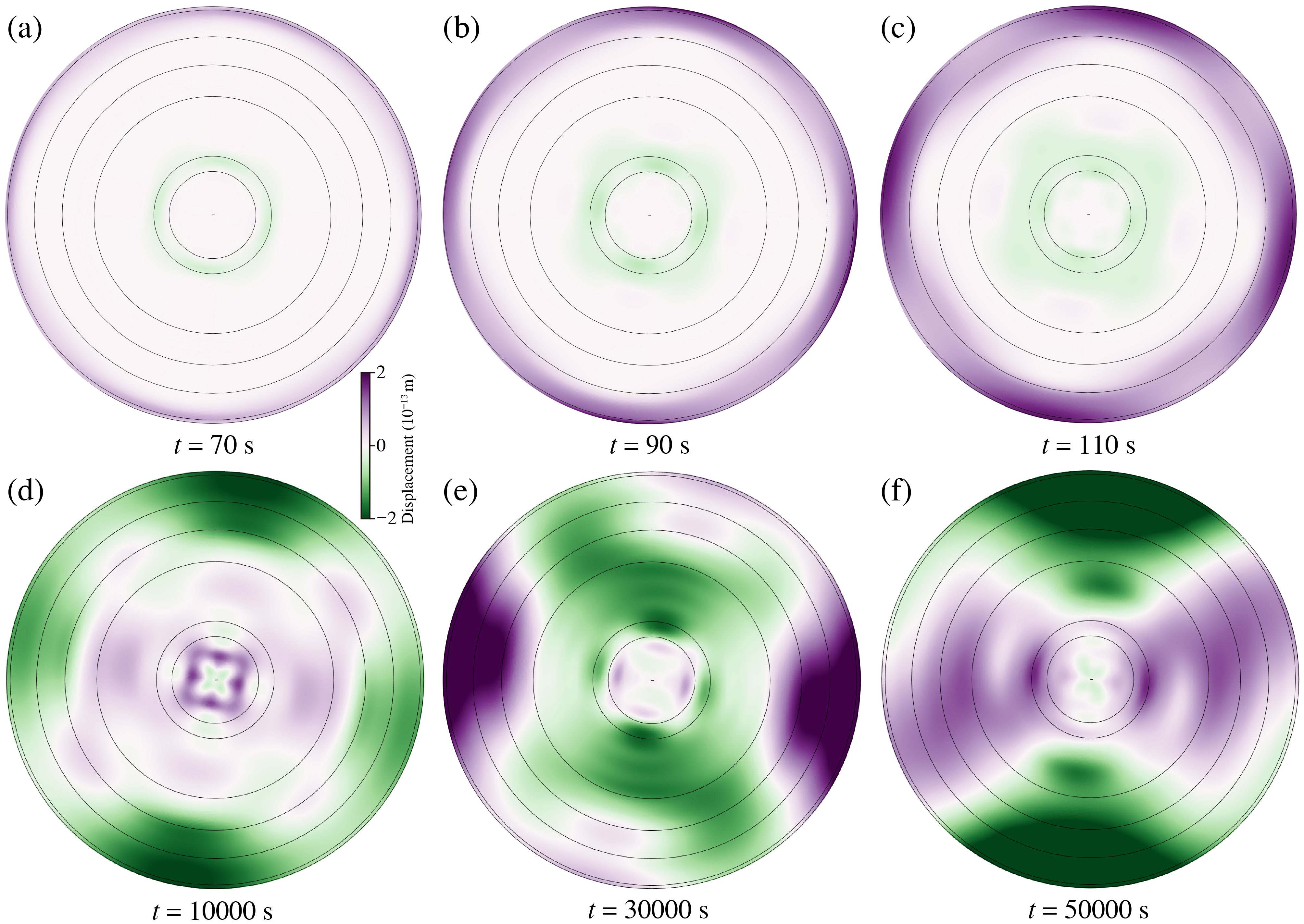}
    \caption{Representative snapshots of the radial displacement of the Moon driven by passing GWs. See the radial displacement movie \textit{Dis\_snapshots.gif} in \cite{webs}. (a) $t = 70~$s; (b) $t = 90~$s; (c) $t = 110~$s; (d) $t = 10,000~$s; (e) $t = 30,000~$s; (f) $t = 50,000~$s.}
    \label{fig:dis}
\end{figure*}
\begin{figure*}[htp]
    \centering
    \includegraphics[width=14cm]{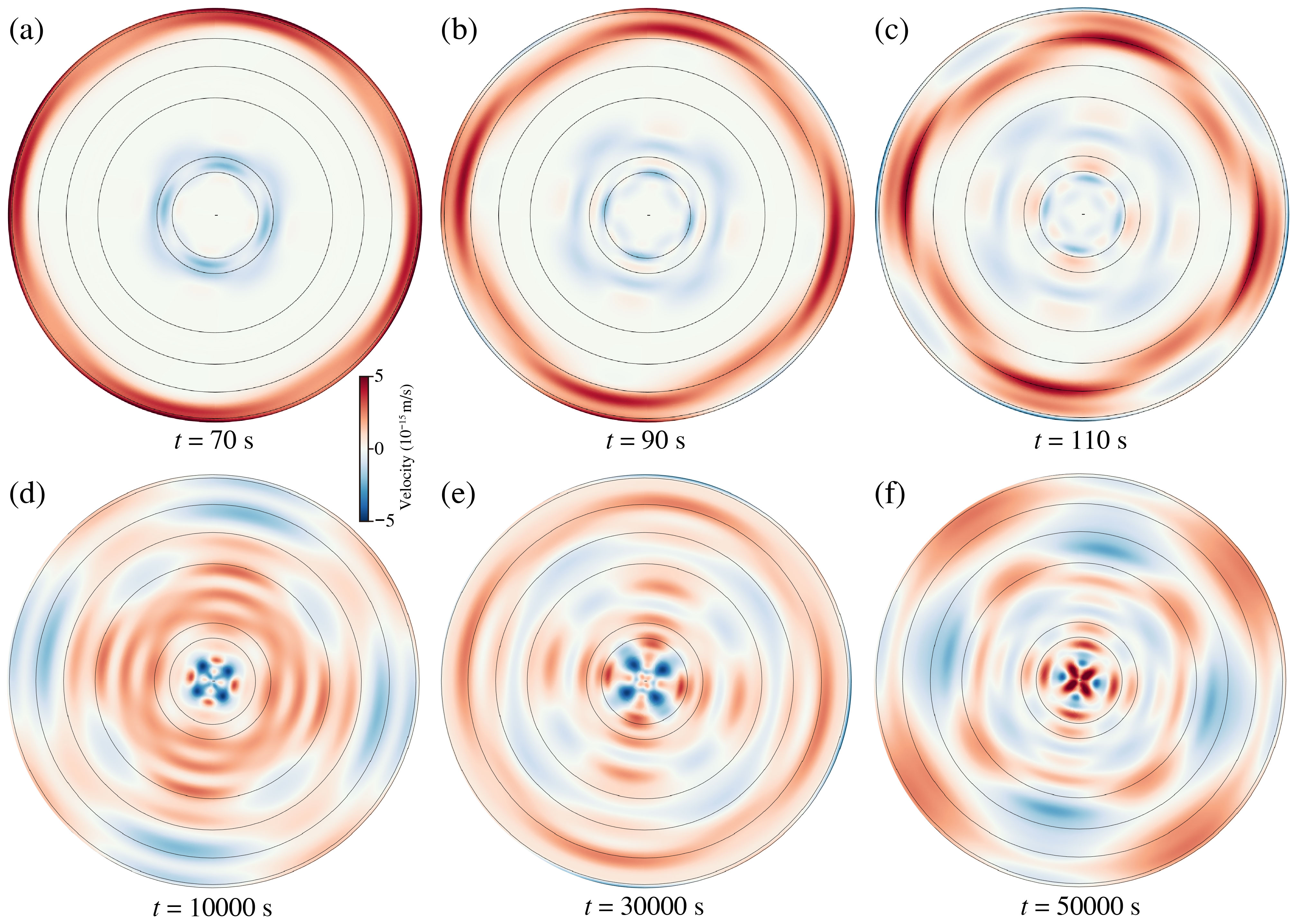}
    \caption{Representative snapshots for the radial velocity. (a) $t = 70~$s; (b) $t = 90~$s; (c) $t = 110~$s; (d) $t = 10,000~$s; (e) $t = 30,000~$s; (f) $t = 50,000~$s.}
    \label{fig:vel}
\end{figure*}

Based on the FEM model (Fig.~\ref{fig:model}), the lunar seismic
response to GW has been simulated. From the radial acceleration snapshots
(Fig.~\ref{fig:dis}), we find that the motion originates at the same location 
where we add the force (Fig.~\ref{fig:dis}a). The response generated
at the core-mantle boundary propagates outward, and meets the response
generated at the lunar crust (Figs.~\ref{fig:dis}\hy{a,b,c}). The seismic waves
interact with each other after meeting, and then spread throughout the Moon
(Fig.~\ref{fig:dis}\hy{d,e,f}), leading to a global oscillation of the Moon. During
this process, normal modes of the Moon are excited and can be recorded on the
surface of the Moon. Fig.~\ref{fig:vel} shows more snapshots of the radial velocity.

\section{\label{sec:DataAnaly}Numerical results of lunar surface response}

\begin{figure*}[htpb]
    \centering
    \includegraphics[width=.95\textwidth]{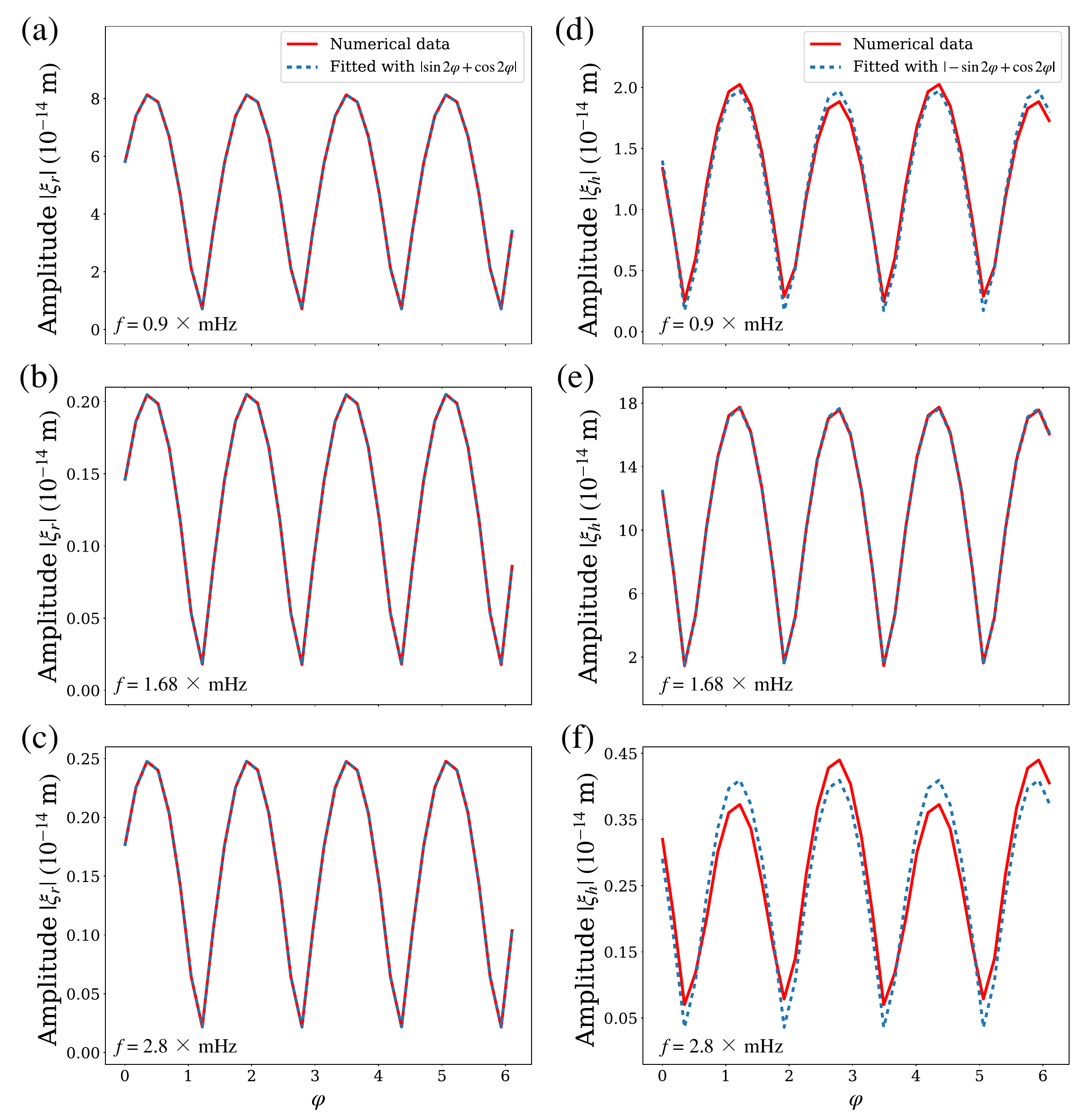}
	\caption{\label{fig:angl}Angular dependence of the radial response $|\xi_{r}(\varphi,f)|$ and horizontal response $|\xi_{h}(\varphi,f)|$. The blue-dashed curves are 
fitting results using Eq.~(\ref{eq:sol_perh_simp}). (a)$\sim$(c) are radial results, (d)$\sim$(f) are horizontal results. The three selected frequencies $f=0.9,~1.68~\text{and}~2.8~$mHz correspond to the first three resonant peaks.}
\end{figure*}

\begin{figure}[htpb]
     \centering
     \includegraphics[width=8cm]{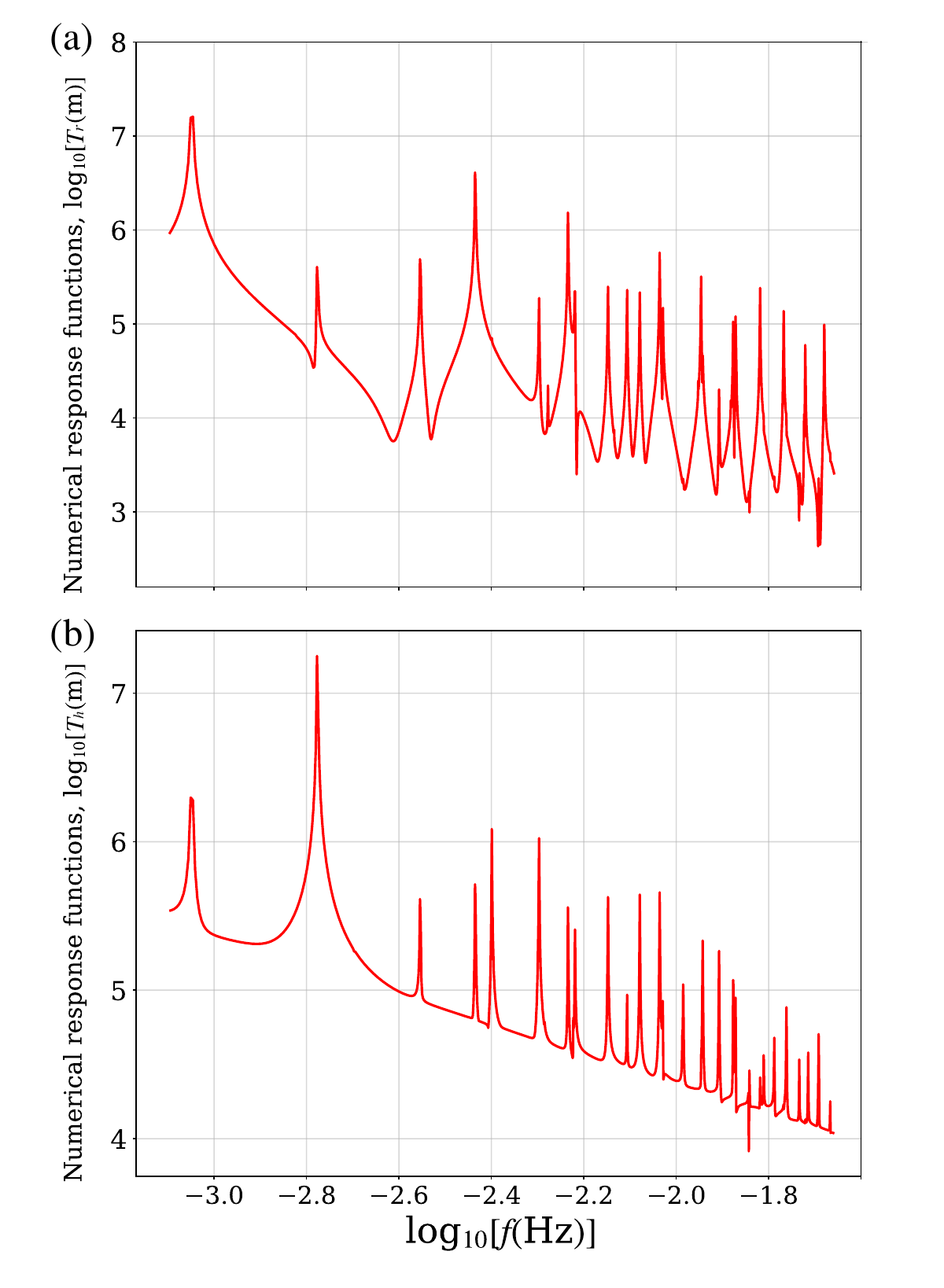}
     \caption{\label{fig:respf_dat}Radial and horizontal response functions (log-scale) per unite strain from our FEM simulations. (a) Radial response function $T_r(f)$. (b) Horizontal response function $T_h(f)$.}
\end{figure}

In this section, we will analyze the main results from our FEM
simulations. We first analyze the angular dependency of the lunar response at different
frequencies. We perform Fourier transform of the lunar surface displacement, 
and obtain the values of $|\xi_{r}(\varphi,f)|$ and
$|\xi_{h}(\varphi,f)|$ for different $\varphi$. To
better compare with \hy{theoretical results}, we select several typical frequencies, $f =
0.9~\text{mHz}$, $1.68~\text{mHz}$ and $2.8~\text{mHz}$, and fit the amplitude
of the lunar surface oscillation assuming an 
angular dependence given by  Eq.~(\ref{eq:sol_perh_simp}). These three selected
frequencies correspond to the first three resonant peaks from FEM simulation
(see Fig.~\ref{fig:respf_dat}), where the response reaches local maximum in the
frequency domain. The results are plotted in Fig.~\ref{fig:angl}.  We find that
the numerical results match generally well with the theoretical relation in
Eq.~(\ref{eq:sol_perh_simp}), but we also see a slight deviation for the horizontal
response. 

Next, we want to directly obtain $T_r(f)$ and $T_h(f)$ from our numerical
results. Our approach is to first calculate
$|\xi_{r}(f,\varphi)/h_{0}g(\omega)|$ and $|\xi_{h}(f,\varphi)/h_{0}g(\omega)|$
for each output position. We eliminate the angular variation by 
averaging over $\varphi$ and then modifying it by a factor of 
$2\sqrt{2}/\pi\simeq 0.9$,
which
comes from the average of $\left | \sin 2\varphi \pm \cos 2\varphi \right |$
[notice another factor of $2$ for the horizontal response, see Eq.~(\ref{eq:sol_perh_simp})].
The results are plotted in Fig.~\ref{fig:respf_dat}. In our calculation we set
$h_{0} = 10^{-20}$, which explains the different amplitudes with respect to
Fig.~\ref{fig:angl}. We will compare this result with the previous semi-analytical
prediction (from Sec.~\ref{sec:theo}) in Sec.~\ref{sec:diss}.

\section{\label{sec:diss}Discussions}

\begin{figure}[htp]
    \centering
    \includegraphics[width=8.5cm]{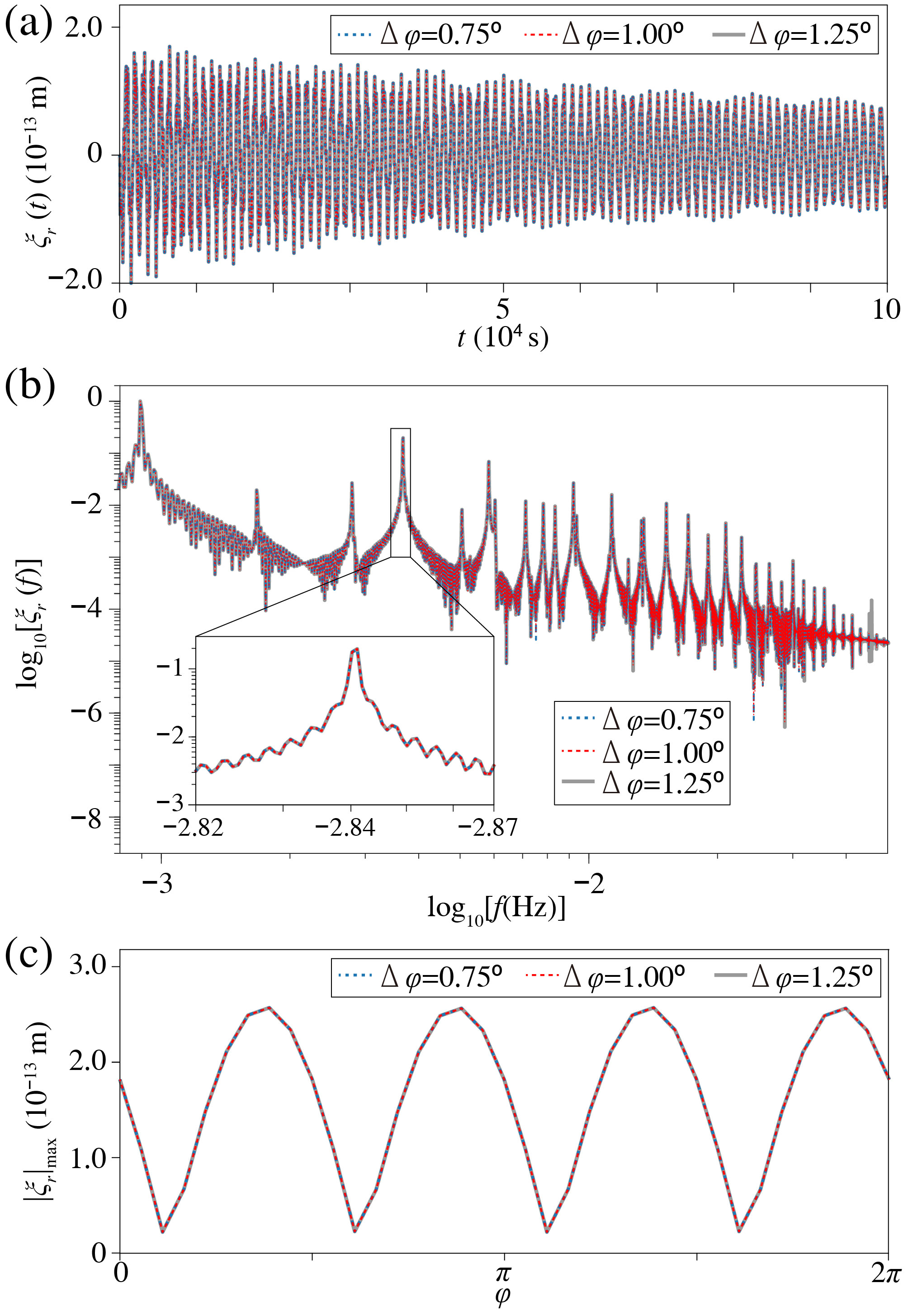}
\caption{Comparing the results derived from three different angular resolutions
of the GW force density, namely, 
$\Delta\varphi = 0.75^{\circ}$, $1.00^{\circ}$, and $1.25^{\circ}~$.
(a) Waveform $\xi_{r}(t)$; (b) normalized spectrum $\xi_{r}(f)$ with the intersecting plot showing
the details at the peak; 
(c) amplitude $|\xi_{r}|_{\text{max}}$ of the waveform as a function of the azimuthal angle 
	$\varphi = 0 \sim 2\pi$. Panels (a) and (b) are derived at $\varphi=0^{\circ}$.}
	\label{fig:dth}
\end{figure}

Since we use FEM to discretize the entire \hy{lunar model}, the locations where we add the forces \hy{may} depend on our choice of the resolution. Such spatial resolution of force density (angular resolution), in principle, should not affect the simulation results if the resolution is sufficiently high. For this reason, we have tested the effects of different angular resolutions (
$\Delta\varphi = 0.75^{\circ}$ , $1.00^{\circ}$, $1.25^{\circ}$) on waveform
and ground amplitude (Fig.~\ref{fig:dth}). The difference in amplitude
among the three is less than $0.1\%$. Therefore, we have chosen 
$\Delta\varphi = 1.00^{\circ}$ as our fiducial resolution. In addition, we would like to emphasize that the
Dyson-type force density (Eq.~\ref{eq:forcedens}) is added at the interface
where the shear modulus and S-wave velocity vary drastically. The
way of constructing the grids and adding the force at these places 
could affect the numerical results of the lunar response to GWs, which is worth
investigating in the future.

\begin{figure}[htp]
     \centering
     \includegraphics[width=8.5cm]{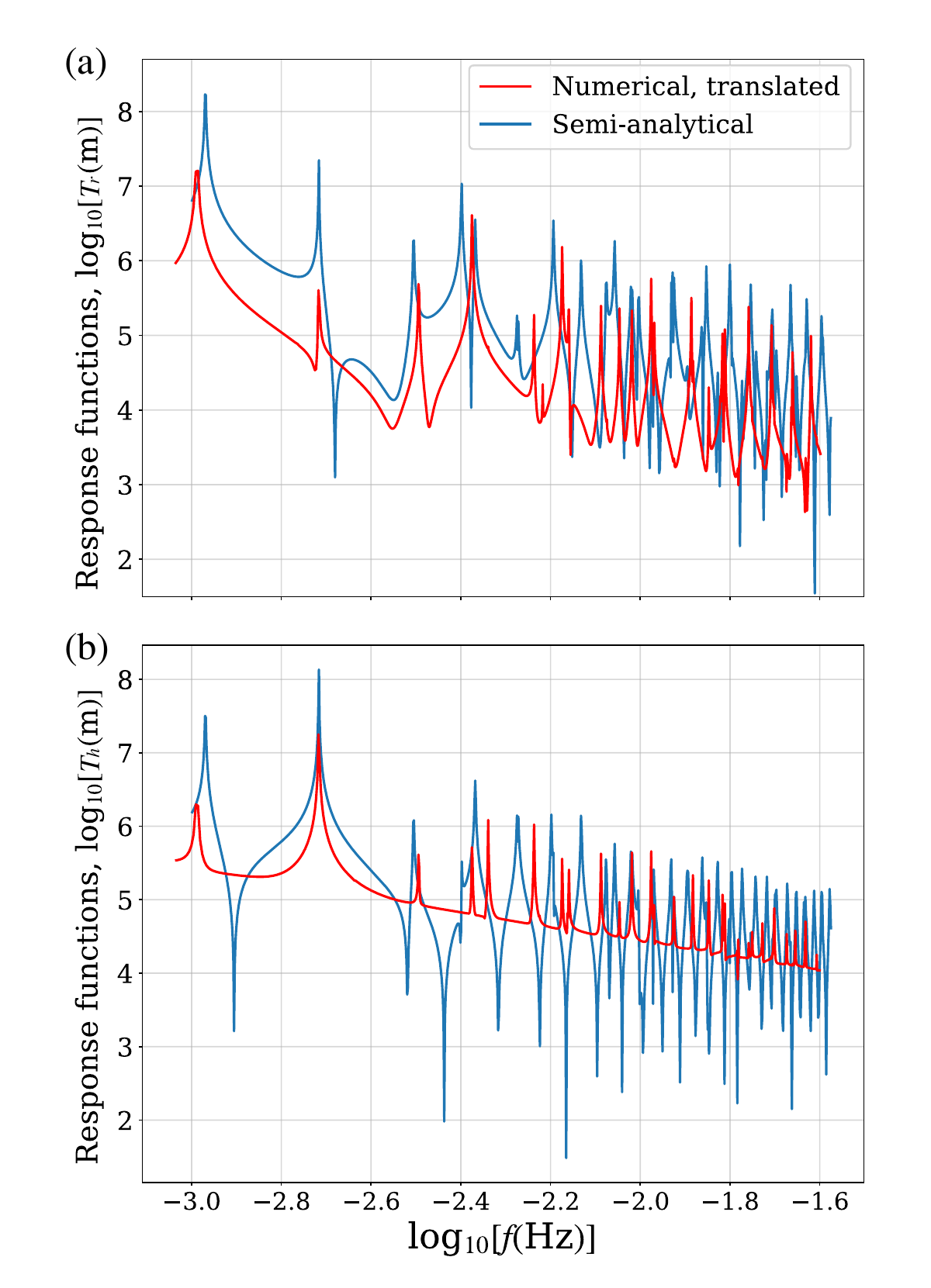}
     \caption{\label{fig:respf_comp}Response function (log-scale) per unite strain from FEM simulation and previous semi-analytical approach. Numerical results have been horizontally translated. (a) Radial response function $T_r(f)$. (b) Horizontal response function $T_h(f)$.}
\end{figure}

To compare the results from our FEM calculation and those from
previous semi-analytical approaches (in Sec.~\ref{sec:theo}), we plot these two
kinds of response functions in Fig.~\ref{fig:respf_comp}. The gray lines are
response functions calculated with the approach described in
Sec.~\ref{sec:theo}, using the same lunar model as in the FEM
calculation. The red lines are the same response functions as in
Fig.~\ref{fig:respf_dat}, but horizontally translated by about $0.06$ dex to
compensate for the difference between 2D (numerical) and 3D (analytical)
models. We find that the two types of response functions match qualitatively well in the frequency range of 1$\sim$20 mHz,
especially regarding the order of magnitude and the locations of the first few
resonant peaks. Although we only simulate lunar response to GW in the frequency bands of 1 mHz to 20 mHz, it does not mean that we are limited to this frequency band. In fact, we can handle higher frequency \zl{bands} (e.g., up to 200 mHz based on the model in this paper). Here for a highly robust and stable simulation, we have only used STF within 20 mHz (Fig.~\ref{fig:stf}). With more computer clusters with larger memory in the future, we can handle a higher frequency band, such as up to 1 Hz.

Nevertheless, in Fig.~\ref{fig:respf_comp} we can still find visible deviations in both the locations of high-frequency resonant peaks and the amplitudes at spectral
troughs. There are several possible reasons that \hy{may} cause these deviations. The main one could be the difference between the 2D (cylindrical) Moon model in the FEM simulation and the 3D (spherical) Moon model in those previous semi-analytical calculations. We give a simple but instructive analysis in Appendix \ref{app:SpheCylin}, which shows
that the difference between 2D and 3D models can lead to a shift of the first few eigen-frequencies around several mHz. The direction of the shift is determined by the details of the structure of the Moon, so that a numerical peak can reside at either the left- or the right-hand side of the analytical peak. \hy{This} simple analysis explains our FEM results qualitatively, and it suggests that the different nature of the mathematical solutions in spherical and cylindrical coordinate, as well as the different types of the boundary conditions, can significantly change the eigen-frequencies and other related results. Since our 2D FEM predicts the locations and amplitudes of the strongest resonant peaks reasonably well, the method will be useful for the design of future seismic GW detection projects.

\hy{Beside, we notice that the horizontal response in FEM simulation behaves differently with radial response, especially at those dips where the responses are much smaller than at the peaks (Fig.~\ref{fig:respf_comp}b). This could be because of the inherent limitations of the 2D plane strain model employed in our study. This 2D simplification could lead to a simulation that works better at characterizing the propagation of radial waves than those of the horizontal waves while oversimplifying the waves perpendicular to the plane of the paper. As a result, this thereby results in a relatively larger error in the horizontal component of the response since the lateral reverberations of surface waves could not be fully considered in the 2D models \cite{smerzini2011comparison}. To mitigate this limitation and enhance the accuracy of our simulations, there is a strong expectation that future advancements will see the integration of 3D numerical simulation techniques. These more sophisticated models hold the promise of overcoming the current constraints by providing a comprehensive representation of wave propagation in all three spatial dimensions. By doing so, they would be able to more accurately capture the intricate interplay between radial and horizontal components of the response, thereby reducing the disparity observed in FEM simulations and offering a more holistic understanding of the underlying physical phenomena. Based on the 3D simulations, the lunar topographic effect and the crustal thickness can be more accurately modeled to depict the seismic response to gravitational waves.}

In the future, to realistically simulate the lunar response to GWs using FEM, a long period of simulation time is needed.  Such a requirement poses a big challenge given the current memory size and computational resources. The problem will be even more severe
when extending our model to 3D. Although we adopt a spherically symmetric uniform numerical model for the purpose of comparison with semi-analytical solutions, we have not considered the actual conditions of the Moon such as topographic surface, fractured subsurface, or lateral variation of the interior structure. However, incorporating these aspects into FEM and/or SEM simulations, as well as extending them to 3D scenarios, poses no technical obstacles, and relevant technologies have just achieved breakthroughs\cite{10.1190/geo2021-0503.1}. Moreover, increasing the time step in FEM/SEM simulations is not straightforward due to the stability condition that must be satisfied (e.g., Courant-Friedrichs-Lewy
condition). Therefore, the choice of time step is closely related to the spatial discretization and the wave propagation speed. In the future, to make 3D simulations feasible, it is crucial to implement specific schemes that \zl{allow} larger time steps (e.g., see
Refs.~\cite{gao2018removing,gao2019extending,miao2023optimal}).

\section{\label{sec:ccl}Conclusion}

In this work, we conduct the first 2D FEM simulation of the lunar response to
GWs. Compared to the previous semi-analytical models based on the normal-mode
formalism, the FEM has great potential of accounting for the rough topography
and the strong lateral heterogeneity of the lunar crust, and/or unique interior
geometric structures. Our results suggest that in the ideal case of a symmetric
Moon model, the response functions derived from FEM and previous
semi-analytical approaches agree generally well, which means that our FEM
simulation is feasible for calculating lunar response to GWs. In the future,
FEM can further conduct 3D global Moon simulations, accounting for the
topographic surface/fractured subsurface/lateral heterogeneous interior
structure of the Moon.

\begin{acknowledgments}
This work is supported by the National Natural Science Foundation of China (Grant Numbers: 42325406, 12473037, 42204178), Key Research Program of the CAS (Grant Numbers: ZDBS-SSW-TLC00104 and KGFZD-145-23-15-2) and Key Research Program of the Institute of Geology and Geophysics, Chinese Academy of Sciences (Grant Numbers: IGGCAS-202203, \hy{IGGCAS-202204 and IGGCAS-202403}).

\end{acknowledgments}

\appendix

\section{\label{app:SpheCylin}Compare the solutions in 3D (spherical) and 2D (cylindrical) models} 

In this appendix, we would like to give a semi-analytical explanation for the
deviation between our numerical results and those previous theoretical ones.
The deviation originates from the fact that in our FEM simulation, we
use a 2D grid which represents an cylindrical ``Moon'' with infinite length.
The radial distribution of this cylinder follows the same radial distribution of 
the 3D spherical Moon. 
To facilitate the comparison, we calculate the
eigen-frequencies of two homogeneous ``Moons'', one is spherical and the other is
cylindrical. These two models have exactly the same physics parameters (radius
$R$, density $\rho$, and two wave velocities $v_{p}$ and $v_{s}$). The results
can therefore be given analytically, as we will show below. Our calculation mainly follows
Ref.~\cite{1995PhRvD..52..591L}, which originally is conducted for a sphere.

As has been proved previously, the spatial part of the eigen-vibrations $\mathbf{u}  \left ( \mathbf{x}   \right )$ with angular frequency $\omega~(=2\pi f)$ can be separated to an irrotational part and a divergence-free part:
\begin{eqnarray}
    \mathbf{u}  \left ( \mathbf{x}   \right ) = C_{0} \nabla \phi\left ( \mathbf{x}   \right ) + C_{1} \mathbf{L} \psi\left ( \mathbf{x}   \right ) + C_{2} \nabla \times \mathbf{L} \psi\left ( \mathbf{x}   \right ) , \label{eq:u_sol}
\end{eqnarray}
where $C_{0}$, $C_{1}$ and $C_{2}$ are three constants, and $\mathbf{L} \equiv  \mathbf{x} \times \nabla$. 
Therefore, the latter two terms are divergence-free. 

The particular forms of $\phi\left ( \mathbf{x}   \right )$ and $\psi\left ( \mathbf{x}   \right )$ in cylindrical coordinates are different from those in spherical coordinates.
In a spherical coordinate system $(r,\theta,\varphi)$, as illustrated in Ref.~\cite{1995PhRvD..52..591L}, we have 
\begin{eqnarray}
    &&\phi^{S}\left ( \mathbf{x}   \right ) = j_{l} \left ( qr \right ) Y_{lm} \left ( \theta,\varphi \right )  ~,\nonumber  \\ &&\psi^{S}  \left ( \mathbf{x}   \right ) = j_{l} \left ( kr \right ) Y_{lm} \left ( \theta,\varphi \right )   ~,
\end{eqnarray}
where $j_{l}$ is $l$-th order spherical Bessel function, $Y_{lm}$ is the spherical harmonic, and two wave numbers are defined as
\begin{eqnarray}
    && q^{2} \equiv \frac{\rho }{\lambda +2\mu } \omega ^{2} ~,\nonumber \\ && k^{2} \equiv \frac{\rho }{\mu } \omega ^{2}   ~.
\end{eqnarray}
In a cylindrical coordinate system $(r,\varphi,z)$, the corresponding forms are
\begin{eqnarray}
    &&\phi^{C}\left ( \mathbf{x}   \right ) = J_{m} \left ( qr \right ) e^{im\varphi } ~,\nonumber  \\ &&\psi^{C}  \left ( \mathbf{x}   \right ) = J_{m} \left ( kr \right ) e^{im\varphi }   ~,
\end{eqnarray}
where $J_{m}$ is $m$-th order Bessel function.

To calculate the value of $\omega$, we apply the free boundary condition $\sigma _{ij} n_{j} = 0$, in which $\sigma _{ij} \equiv \frac{1}{2} \left ( u_{i,j} + u_{j,i} \right ) $ and $n_{j}$ is the surface normal vector. For GWs, we only need to consider the spheroidal mode \cite{1996CQGra..13.2865B}, which means $C_{1}=0$. In spherical coordinate, the result has been derived in Ref.~\cite{1995PhRvD..52..591L} which is
\begin{widetext}
\begin{eqnarray}
        \operatorname{det}\left(\begin{array}{cc}
\beta^{S}_{2}(q R)-\frac{\lambda}{2 \mu} q^{2} R^{2} \beta^{S}_{0}(q R) & l(l+1) \beta^{S}_{1}(k R) \\
\beta^{S}_{1}(q R) & \frac{1}{2} \beta^{S}_{2}(k R)+\left[\frac{l(l+1)}{2}-1\right]\beta^{S}_{0}(k R) 
\end{array}\right)=0 ~,
\end{eqnarray}
where
\begin{eqnarray}
        \beta^{S} _{0} (x) \equiv \frac{j_{l} (x)}{x^{2} } , \quad \beta^{S} _{1} (x) \equiv \frac{\mathrm{d} }{\mathrm{d} x} \left [ \frac{j_{l}(x) }{x}  \right ], \quad  \beta^{S} _{2} (x) \equiv \frac{\mathrm{d}^{2}  }{\mathrm{d} x^{2} }\left [ j_{l} (x) \right ]  ~.
\end{eqnarray}

In cylindrical coordinate, after some algebra, we find that 
\begin{eqnarray}
         \operatorname{det}\left(\begin{array}{cc}
\beta^{C}_{2}(q R)-\frac{\lambda}{2 \mu} \beta^{C}_{0}(q R) & \beta^{C}_{2}(k R)+m^{2} \beta^{C}_{1}(k R) \\
2 \beta^{C}_{1}(q R)+\frac{\beta^{C}_{0}(q R)}{q^{2} R^{2}} & \beta^{C}_{2}(k R)+2 \beta^{C}_{1}(k R)+\frac{m^{2}+1}{k^{2} R^{2}} \beta^{C}_{0}(k R)
\end{array}\right)=0 ~,
\end{eqnarray}
where
\begin{eqnarray}
        \beta^{C} _{0} (x) \equiv J_{m} (x), \quad \beta^{C} _{1} (x) \equiv \frac{\mathrm{d} }{\mathrm{d} x} \left [ \frac{J_{m}(x) }{x}  \right ], \quad  \beta^{C} _{2} (x) \equiv \frac{\mathrm{d}^{2}  }{\mathrm{d} x^{2} }\left [ J_{m} (x) \right ]  ~.
\end{eqnarray}
\end{widetext}

For GWs, it is required that $l=2$ in spherical coordinate. However, 
in cylindrical coordinate the choice of $m$ is unclear so far
(we will investigate this issue in the future), so we calculate both $m=2$ and $m=2.5$. 
The first one is the same as the $l$ index in the spherical harmonic, 
and the second one comes from the mathematical relation between the Bessel function and the 
spherical Bessel function:
\begin{eqnarray}
    j_{l} \left ( x \right )  = \sqrt{\frac{\pi }{2x} } J_{l+0.5} \left ( x \right ) ,~l> 0  \nonumber ~.
\end{eqnarray}
We calculate the first few eigen-frequencies $f = \omega/2\pi$ for a
homogeneous ``Moon'' with $R = 1737~\text{km}$, $\rho =
3\times10^{3}~\text{kg/m}^{3}$, $v_{p} = 8\times10^{3}~\text{m/s}$ and $v_{s} =
4\times10^{3}~\text{m/s}$
\cite{2011Sci...331..309W,2011PEPI..188...96G,2023Natur.617..743B} [notice that
$\lambda = \rho(v_{p}^{2}-2v_{s}^{2})$ and $\mu = \rho v_{s}^{2}$]. The results
are listed in Table~\ref{tab:eigf}. The frequencies in the $m=2$ case 
are more similar to
our numerical results. Since these frequencies are derived from a different lunar model (one is
homogeneous and the other is heterogeneous), the comparison is 
not exact. We would also like to point out that, in principle, we can write down the
complete solutions based on Eq.~(\ref{eq:u_sol}) and the particular form of the
force density acting on the Moon. Nevertheless, the current comparison, also not exact, 
already qualitatively explains the deviation between our semi-analytical and numerical
results.

\begin{table}[H]
\caption{\label{tab:eigf}
First few eigen-frequencies for a 3D spherical ``Moon'' and a 2D cylindrical ``Moon'' with the same parameters.}
\begin{ruledtabular}
\begin{tabular}{cccccc}
 &Mode&$f$ (mHz)& &Mode&$f$ (mHz) \\
\hline
Spherical&$l=2$& 0.971 & Cylindrical & $m=2$ & 0.666  \\
& & 1.868 &   &   & 1.970 \\
& & 3.161 &   &   & 3.087 \\
& & 4.026 &   &   & 3.508 \\
& & 4.505 &   &   & 4.300 \\
& & ... &   &   & ... \\
& &  &   & $m=2.5$  & 0.917 \\
& &  &   &  & 2.217 \\
& &  &   &  & 3.330 \\
& &  &   &  & 3.966 \\
& &  &   &  & 4.575 \\
& &  &   &  & ... \\
\end{tabular}
\end{ruledtabular}
\end{table}
 
\bibliography{main}

\end{document}